\documentclass[aps,superscriptaddress,longbibliography,preprint]{revtex4-1}
\usepackage{amsmath,amssymb,pifont,pxfonts,fdsymbol}
\usepackage{graphicx,color,subfigure,bm}
\usepackage{multirow}
\usepackage[usenames,dvipsnames]{pstricks}
\usepackage{epsfig}
\usepackage[colorlinks=true, citecolor=red, linkcolor=blue,urlcolor=blue ]{hyperref}
\graphicspath{{img}}

\def\De{{\rm De}}
\def\Wi{{\rm Wi}}

\begin{document}


\title{Front-back asymmetry controls the impact of viscoelasticity on helical swimming}%

\author{V. Angeles}
\affiliation{Instituto de Investigaciones en Materiales, Universidad Nacional Aut\'onoma de M\'exico, Apdo. Postal 70-360, M\'exico Distrito Federal 04510, M\'exico}

\author{F.A. God\'inez}
\affiliation{Instituto de Ingenieria, Universidad Nacional Aut\'onoma de M\'exico, Apdo. Postal 70-360, M\'exico Distrito Federal 04510, M\'exico}
\affiliation{Polo Universitario de Tecnolog\'ia Avanzada, Universidad
Nacional Aut\'onoma de M\'exico, Apodaca 66629,
Nuevo Leon, M\'exico}

\author{J.A. Puente-Velazquez}
\affiliation{Instituto de Investigaciones en Materiales, Universidad Nacional Aut\'onoma de M\'exico, Apdo. Postal 70-360, M\'exico Distrito Federal 04510, M\'exico}

\author{R. Mendez}
\affiliation{Instituto de Investigaciones en Materiales, Universidad Nacional Aut\'onoma de M\'exico, Apdo. Postal 70-360, M\'exico Distrito Federal 04510, M\'exico}

\author{E. Lauga}
\email{e.lauga@damtp.cam.ac.uk}
\affiliation{Department of Applied Mathematics and Theoretical Physics,
University of Cambridge, Cambridge CB3~0WA, United Kingdom}

\author{R. Zenit}
\email{zenit@brown.edu}
\affiliation{Center for Fluid Mechanics, School of Engineering, Brown University, Providence RI 02912, USA}
\affiliation{Instituto de Investigaciones en Materiales, Universidad Nacional Aut\'onoma de M\'exico, Apdo. Postal 70-360, M\'exico Distrito Federal 04510, M\'exico}

\date{\today}
\begin{abstract}
We conduct experiments with force-free magnetically-driven rigid helical swimmers in Newtonian and viscoelastic (Boger) fluids. 
By varying the  sizes of the swimmer body and its helical tail, we show that the impact of viscoelasticity strongly   depends on the swimmer geometry: it can lead to a significant increase of  the swimming speed  (up to a factor of five), a similar decrease  (also up to a factor of five) or it can have  approximately no impact. Analysis of our data along with theoretical modeling shows that the influence of  viscoelasticity on helical propulsion is  controlled by a snowman-like effect, previously reported for dumbbell swimmers, wherein the front-back asymmetry of the swimmer leads to a non-Newtonian elastic force that can either  favor or hinder locomotion.
\bigskip


\end{abstract}

\maketitle

\section{Introduction}

Microorganisms swim in an environment in which inertial effects are negligible~\cite{Powers2009} and therefore they employ locomotion strategies  very different from those of fish and humans. The scallop theorem~\cite{Purcell1997}, that states that a simply-articulated time-reversible swimmer cannot achieve locomotion in a Stokes flow, provides a  clear illustration of the implications of living in a  viscous-dominated  environment. 

There are several methods exploited by microorganisms to cope with   environments dominated by viscosity. In particular, the majority of motile bacteria, simple single-cell organisms,   exploit helical  flagellar filaments in order to achieve locomotion~\cite{Lauga2016}. These  semi-rigid filaments can either be used in isolation (monotrichous  bacteria) or, for cells with several helical filaments (peritrichous bacteria), they can  bundle together to form a single helical structure. In all cases, propulsion of the cell is enabled by the rotation of  semi-rigid helical filaments in the viscous fluid.  
 Since a helix is a chiral shape, a rotation around the helical axis bypasses the constraints of the scallop theorem and it is able to generate viscous thrust along its axis.

The mechanics of helical swimming is  well understood  in the case of Newtonian flows~\cite{Rodenborn2013}. However  many of the fluids in which microorganisms move are not Newtonian, ranging from mucus  and complex suspensions to  biological tissues. As with most flows in which such fluids are involved, the   dynamics of swimming microorganisms is significantly affected by viscoelasticity, the presence of shear-dependent stresses, or both. 
Numerous  studies have recently been devoted to the subject~\cite{Lauga2007,liu2011,Arratia2011,Arratia2013,Zenit2013,Kudrolli2013, spagnolie2013,Becca2014,Godinez2015}, with some results which appear to be in  contradiction with each other, and thus a number of   fundamental issues remain.

One possible starting point to capture the effect of viscoelasticity is the theoretical study in Ref.~\cite{Lauga2007} which  extended   the classical Taylor swimming sheet result to  the case of viscoelastic Oldroyd-like fluids. The swimming speed of the sheet, $U_{NN}$, normalized  by its Newtonian value, $U_{N}$, was calculated at leading order in the waving amplitude to be
\begin{equation}\label{eqn:lauga2007}
    \frac{U_{NN}}{U_N}= \frac{1+\kappa \De^2}{1+\De^2},
\end{equation}
where $\kappa=\mu_S/\mu \leq 1$ is the ratio of the solvent to total viscosity. Here $\De=  {\tau} \omega$ is the Deborah number,  
where $\tau$ is the fluid relaxation time and   $\omega$ the angular frequency of the wave,    a dimensionless parameter measuring the relative importance of viscoelasticity in a given flow. Since $\kappa\leq 1$, the result in Eq.~\eqref{eqn:lauga2007} predicts that the swimming speed in a viscoelastic fluid will be smaller that its Newtonian equivalent for any value of $\De$. While this result reignited  interest in the field, its validity is restricted to the case of small wave amplitude ($ak<<1$, where $a$ and $k$ are the amplitude and wave number of the oscillation) and to the case in which the wave is not affected by the nature of the surrounding fluid (fixed-kinematics). In that limit,  both experiments~\cite{Arratia2011} and numerical simulations~\cite{spagnolie2013} have shown that this prediction is correct.

In contrast with the result above, numerical computations found that when the amplitude of oscillation was not small, the swimming speed in the  viscoelastic fluid could be larger than that in the Newtonian fluid~\cite{Shelley2010}. Several experimental reports have subsequently confirmed that a faster speed in viscoelastic media was in fact possible~\cite{liu2011,Zenit2013}. The possibility of obtaining both decrease and increase in swimming was  reported in Ref.~\cite{Godinez2015} where experimental measurements for the ratio of swimming speeds for three different swimming strategies at fixed $\De$ number showed that the swimming ratio could be smaller, larger or approximately one depending on the swimming kinematics.
In other words, the swimming speed in a viscoelastic fluid does not depend solely on the value of the $\De$ number. A recent analysis of the effect of the swimming gait on   locomotion in non-Newtonian media obtained theoretical predictions   in good agreement with experiments so far~\cite{Elfring2016}.

Given the complexity that arises from having swimming  in which the waving shape of the  appendages might depend on the flow itself via mechanical feedback, it is simpler to focus first  on the case  for which the kinematics are fixed. The biological example where the shape is known to be essentially rigid and unchanged by the  the fluids is the rotating helical filaments of swimming bacteria. The work in Refs.~\cite{Wolgemuth2007,Fu2009}
 extended the Taylor swimming sheet result from Ref.~\cite{Lauga2007} to the case of helix in the limiting case of a small pitch angle $\theta$ (i.e.~the angle between the helix axis and the local tangent along the helix centrerline). They obtained  the same decreasing trend of the normalized  swimming speed with Deborah number as in Eq.~\eqref{eqn:lauga2007}. Subsequent  experiments  with   force-free helices driven in rotation showed, in contrast, that the helical swimming speed could,   be larger than that in the Newtonian case~\cite{liu2011}. Specifically,    the swimming speed was shown to  depend on  both the value of the Deborah number and the shape of the helix and helices with larger pitch angle produced more pronounced increase in swimming. However, only two values of the pitch angles were tested experimentally~\cite{liu2011}.  Subsequent numerical simulations confirmed that the normalized swimming speed could be smaller or larger than one, depending on both Deborah number and the geometry of the helix~\cite{spagnolie2013}. Related   work   showed that the drag force on slender cylinders in viscoelastic fluids --  the required building blocks to understand force generation  for rotating helices -- depend strongly  on their orientation relative to the main flow and their drag experiences strong tip  effects~\cite{Li2017,Li2018}.  Recent experimental measurements using live bacteria  showed that shear thinning effects lead to higher swimming speeds that those with Newtonian fluids~\cite{Qu2020}, in agreement with past work~\cite{li15,gomez16}; however, the fluid viscoelasticity did not affect the swimming speed directly but instead  the unsteady bundling/unbundling dynamics of the bacteria flagella.

It is therefore clear that, in addition to the expected dependence on the value of the Deborah number, the geometrical properties of a helix   impact its free swimming speed in a non-Newtonian fluid.  In this paper, we conduct experiments with force-free magnetically-driven rigid helical swimmers in Newtonian and viscoelastic (Boger) fluids. We measure the swimming speeds for helices  with many different  geometries and relative head sizes. In accordance with previous studies, we found that depending on the helical geometry their swimming speeds  can either increase significantly (up to a factor of five), decrease (also up to a factor of five) or remain approximately unchanged.  The increase vs decrease of the  normalized swimming speed for all of our experimental results appears to be correlated to the front-back asymmetry in size: when the helix has a larger diameter that the head, a swimming speed larger than the Newtonian value is observed, and vice-versa.  The impact of viscoelasticity on helical swimming appears  thus to be controlled by the snowman effect,  proposed theoretically~\cite{pak12} and corroborated experimentally~\cite{puente2019} in past work, wherein an elastic force driven by normal stress differences is generated in the  viscoelastic fluid by the rotation of the swimmer. Adapting the modeling from Ref.~\cite{pak12}, we show, in agreement with our experiments, that this elastic force can then either hinder or favor propulsion depending on the ration between the size of the swimmer's body and that of its helical tail. 
 
 \section{Experimental setup}
\begin{figure}
\noindent\includegraphics[width=0.5\textwidth]{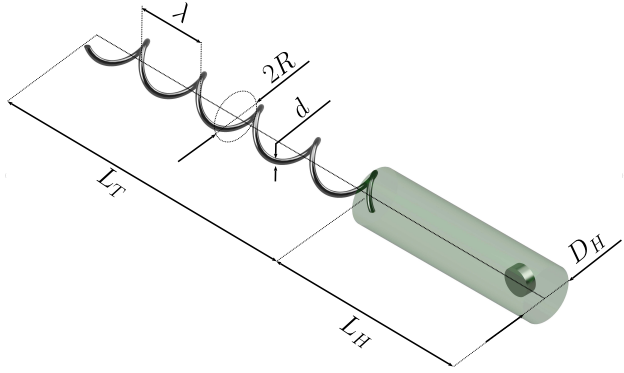}
\caption{Sketch of the design parameters of the   magnetically-driven rigid helical swimmers. For the helix,  $R$ is the radius, $\lambda$ is the  wavelength (pitch), $L_{T}$ is the projected length, $d$ is the diameter of the helical filament. For the head, $L_{H}$ is the length   and $D_{H} $ is the diameter.}
\label{Fig1}
\end{figure}

The experimental design is similar to that previously used in Ref.~\cite{gomez16}.  
A force-free swimmer consisting of a tubular plastic head with a rigid helix tail is placed inside a test fluid. By inserting a small permanent magnet inside the head, the swimmers can be rotated under the action of an external rotating magnetic field~\cite{Godinez2012}.   The shape of the swimmers is depicted schematically in Fig.~\ref{Fig1}. A right-handed rigid helix was placed at the other end of the cylindrical head. In all cases, both the size of the head (length $L_H$ and diameter $D_H$) and the helix (contour length $L$, projected length $L_T $, radius $R$, wavelength $\lambda$ and filament diameter $d$) were varied in order to explore the effect of geometry as widely as possible. The values of the geometrical parameters for all swimmers used in this study are shown in Table~\ref{Table1}. The first five swimmers (F1 and R1 to R4) had tails made of steel wire (Young's modulus $E \approx$ 207 GPa). The second set of swimmers (A1 to A5) were 3D printed and the tail  fabricated with a polymeric resin. Note that the pitch angle of the helix, $\theta$,   defined as
$\tan \theta = {2\pi R}/{\lambda}
$,   varies in our experiments  from 29$^\circ$ to 77$^\circ$.

\begin{table*}[t]
\centering
\begin{tabular}{c|c|c|c|c|c|c|c|c|c|c}
\hspace{2mm} {\bf Swimmer} \hspace{2mm} & \hspace{2mm} {$L_H$}\hspace{2mm}  & \hspace{2mm} {$D_H$}\hspace{2mm}  & \hspace{2mm} {$L$}\hspace{2mm}  & \hspace{2mm} {$d$}\hspace{2mm}  & \hspace{2mm} {$\lambda$} \hspace{2mm} & \hspace{2mm} {$2R$}\hspace{2mm}  & \hspace{2mm} $R/\lambda$ \hspace{2mm} & \hspace{2mm} {$\theta$} \hspace{2mm} & \hspace{2mm} $D^*=2R/D_H$ \hspace{2mm} & \hspace{2mm} {\,\,$L_T$\,\,}\hspace{2mm}  \\ \hline
F1  ({\color[rgb]{0.5,0.5,0.5}$\triangle$,$\blacktriangle$})     & 14.3                      & 4.0                       & 58                      & 0.3                     & 7.6                          & 3.5                      & 0.23 &\hspace{2mm} 0.97 (55$^\circ$)     \hspace{2mm}               & 0.88 & 35.7            \\
R1 ({\color[rgb]{1,0,0}$\medcirc$, $\medbullet$})     & 23                        & 3.0                       & 65                      & 0.9                     & 10.0                          & 1.8                      & 0.09 & 0.52 (29$^\circ$)                    & 0.60 & 56.8                      \\
R2 ({\color[rgb]{1,0.5,0}$\Box$, $\blacksquare$})     & 23                        & 3.0                       & 65                      & 0.9                     & 10.0                          & 3.2     & 0.16  & 0.79 (45$^\circ$)                    & 1.07 & 45.9                      \\
R3 ({\color[rgb]{1.0,0.85,0}$\triangledown$,$\blacktriangledown$})     & 23                        & 3.0                       & 65                      & 0.9                     & 10.0                          & 4.6                      & 0.23  & 0.96 (55$^\circ$)                    & 1.53 & 37.3                      \\
R4 ({\color[rgb]{0,1,0}$\lozenge$,$\blacklozenge$})     & 23                        & 3.0                       & 65                      & 0.9                     & 10.0                          & 11.8                     &  0.59 & 1.31 (75$^\circ$)                    & 3.93 &16.8                      \\
A1 ({\color[rgb]{0.5,0,1}$\triangleleft$,$\blacktriangleleft$})     & 17.3                      & 4.1                       & 76                      & 1.0                     & 9.5                           & 3.0                      & 0.16 & 0.75 (45$^\circ$)                    & 0.73 & 54.3                      \\
A2 ({\color[rgb]{0,0,1}$\medtriangleright$,$\blacktriangleright$})    & 17.3                      & 4.1                       & 80                      & 1.0                     & 9.5                           & 7.0                      & 0.37 & 1.13 (67$^\circ$)                    & 1.71 & 39.36                     \\
A3 ({\color[rgb]{0,1,1}$\triangle$,$\blacktriangle$})    & 17.3                      & 4.1                       & 83                      & 1.0                     & 9.5                           & 15.0                     & 0.79 &1.34 (77$^\circ$)                    & 3.66 & 23.2                      \\
A4   ({\color[rgb]{0,0,0}$\medcirc$, $\medbullet$})   & 17.3                      & 4.1                       & 80                      & 1.0                     & 5.0                           & 3.5                      & 0.35 & 1.13 (65$^\circ$)                    & 0.85 & 37.6                      \\
A5  ({\color[rgb]{1,0,1}$\Box$, $\blacksquare$})     & 17.3                      & 4.1                       & 80                      & 1.0                     & 13.0                          & 9.0                      & 0.35 & 1.13 (65$^\circ$)                    & 2.20 & 37.2
\end{tabular}
\caption{Dimensions of the ten helical swimmers used in this study. All length are reported in millimeters. Symbols are defined  in Fig.~\ref{Fig1} while $L$ is the total contour length of the tail and $\theta$ is the pitch angle in radians (degrees). The empty and solid symbols represent experiments conducted in Newtonian and Boger fluids, respectively.}
\label{Table1}
\end{table*}

The rotation of the head, when combined with a chiral tail shape, produces the thrust force that propels the swimmer through the fluid. The rotation frequency of the external magnetic field,  measured with a digital tachometer,   ranged from 0.41~Hz to 5.8~Hz, with a different  range  for each swimmer. All experiments are conducted below the step-out frequency when the swimming no longer rotates with the external   frequency.  

Two types of fluids were used, a Newtonian and a viscoelastic Boger fluid, and in both cases we have two test fluids. 
The viscoelastic fluid was prepared with glucose, water and a small amount of polyacrylamide (PAA, molecular weight $5\times10^6$ g/mol from Sigma-Aldrich). The Boger fluids were was fabricated by slowly dissolving the  polyacrylamide in non-ionic water for 24 hours. Afterwards, the polymeric solution was added to the glucose and the mixture was mixed slowly for four days. We show in 
Table~\ref{Table2}  the properties of the two pairs of tests fluids used in this study.  For the first test pair, we used industrial grade glucose and for the second commercial corn syrup (Karo brand). While the properties of the industrial grade glucose varied from batch to batch, commercial corn syrup was consistently the same. In both cases, the Newtonian reference liquid was fabricated by adding water to glucose until the fluid had similar viscosity to that of the viscoelastic fluid. 

\begin{table}[b]
\centering 
\begin{tabular}{c |c| c| c| c| c}
{\bf Fluid}  & $G/W/PAAM$ & $\rho $ & $\mu $ & $n$ & $\tau $ \\
& (\%) &  \hspace{2mm}kg/m$^3$\hspace{2mm} & \hspace{2mm}Pa\,s\hspace{2mm} & (-) & \hspace{2mm}s\hspace{2mm} \\
\hline
N1  & 89/11/0 & 1390 & 3.5 & 1.0 & 0.0\\[0.5ex]
B1 & \hspace{2mm}84.96/15/0.04\hspace{2mm} & 1340 & 3.8 & 0.98 & 1.23\\
N2 & 89/11/0 & 1385 & 1.64 & 1.0 & 0.0\\[0.5ex]
B2 & 84.96/15/0.04 & 1366 & 1.64 & \hspace{2mm}0.98 \hspace{2mm}& \hspace{2mm}1.63\hspace{2mm}
\end{tabular}
\caption{Composition and physical properties of the four fluids studied: Composition [from glucose (G), water (W) and polyacrylamide (PAAM)], mass density ($\rho$), dynamic viscosity ($\mu$), power index ($n$), and mean relaxation time ($\tau$).} 
\label{Table2}
\end{table}

The rheological properties of the fluids were determined using a rheometer with parallel plates with 40~mm   diameter and 1~mm gap (TA Instruments, ARES-G2). Both steady and oscillatory tests were conducted to measure the dynamic viscosity, $\mu$, the storage and loss moduli, $G'$ and $G''$, respectively. The mean relaxation time is calculated by fitting $G'$ and $G''$ to a generalized Maxwell model~\cite{Zenit2013}. The density of the fluids are obtained using a 25~ml pycnometer.

The motion of the swimmer in both Newtonian and a viscoelastic (Boger) fluid  was filmed with a digital camera at 60 frames per second. The images were processed digitally with the software {\it Tracker}. Each experiment was repeated at least three times to ensure repeatability. 
The temperature  in the experiment ranged between $23\,^{\circ}\mathrm{C}$  and $24\,^{\circ}\mathrm{C}$. 

\section{Experimental results}

\begin{figure}[t]
\noindent\includegraphics[width=0.5\textwidth]{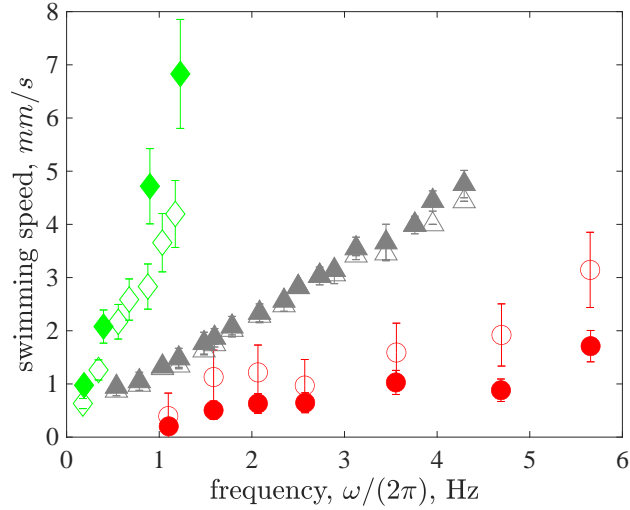}
\caption{Measured swimming speed, $U$,  as a function of rotational frequency, $\omega/2\pi$, for three representative swimmers (F1, R1 and R4 from 
Table~\ref{Table1} where symbols are defined). Empty and filled symbols show the results for Newtonian  and viscoelastic fluids, respectively (fluids N1 and B1 from Table~\ref{Table2}).}
\label{results_1}
\end{figure}

Each swimmer was tested in a fluid pair and its  swimming speed was measured as a function of rotational frequency allowing direct comparison between Newtonian and viscoelastic results. We show in Fig.~\ref{results_1} shows three typical experimental results chosen to illustrate the three possible qualitative results. The swimming speed is plotted as a function of rotational frequency for swimmers F1, R1 and R4 from Table~\ref{Table1} for the first fluid pair N1 and B1, from Table~\ref{Table2}. 
Clearly,  for a helical swimmer,   three different behaviors are possible:  the swimmer can swim faster in a viscoelastic fluid compared to the Newtonian case (green rhombus), is can go slower (red circles) or  with approximately the same speed (grey triangles). The three swimmers, despite the changes in their geometrical parameters, are propelled by the same helical action and the most notable difference between them  are the value of their pitch angle, $\theta$, and tail-to-head size ratios, $D^*=2R/D_H$. The angles range from 29$^\circ$ (slower swimming) to 52$^\circ$ (same speed) to 75$^\circ$ (faster swimming) while the size ratios are $D^*= 0.6$ 
(slower swimming),   0.88 (same speed) and 3.93 (faster swimming). 

\begin{figure}[t]
\noindent\includegraphics[width=0.45\textwidth]{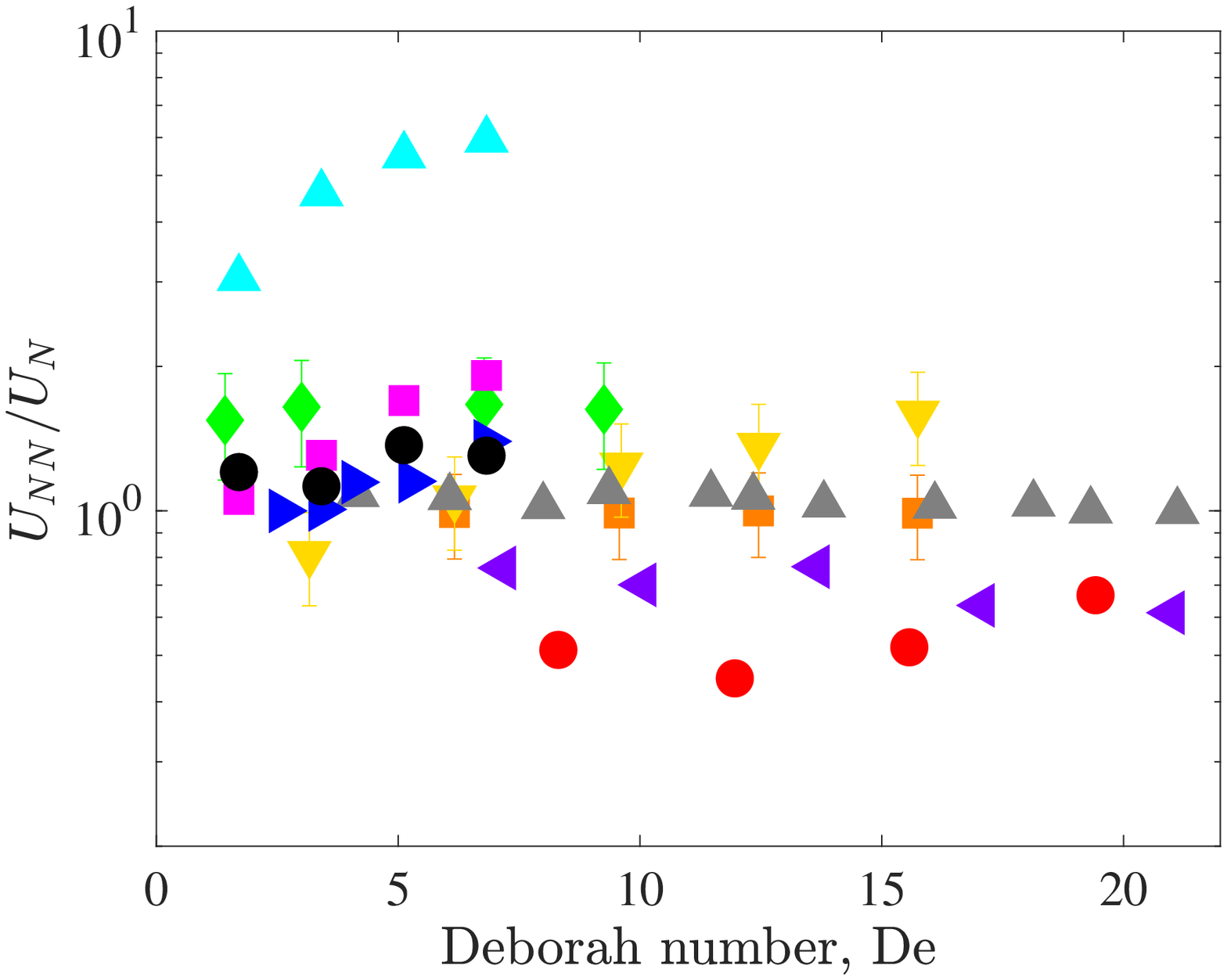}\qquad
\quad\quad\includegraphics[width=0.45\textwidth]{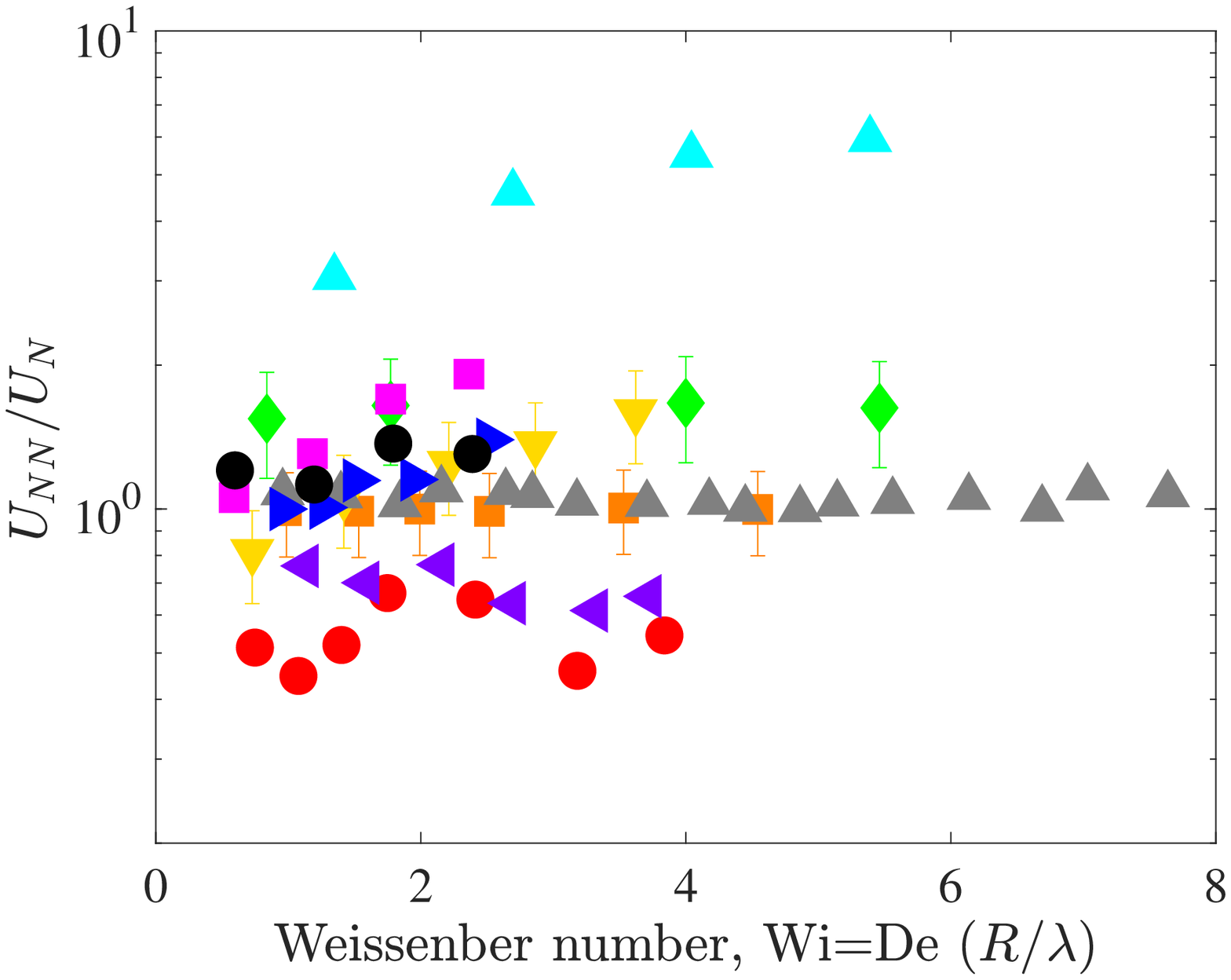}
\caption{Ratio of viscoelastic to Newtonian swimming
speeds, $U_{NN}/U_N$, as a function of the Deborah number (left) and Weissenberg numbers (right).  For symbols, see   Table~\ref{Table1}.}
\label{results_2}
\end{figure}

To quantify the influence of viscoelasticity on the locomotion,  we next calculate the  ratio of the swimming speeds, $U_{NN}/U_N$, 
where $U_{NN}$ and $U_N$ are  the measured mean speeds in the viscoelastic and Newtonian  fluids, respectively. To assess the relative importance of viscoelastic effects, we calculate the Deborah number as $\De =  \omega \tau$, 
where $\tau$ is the fluid relaxation time (from Table~\ref{Table2}). 
The ratio  $U_{NN}/U_N$ is then plotted in Fig.~\ref{results_2} (left)    as a function of $\De$ for all the swimmers studied here (from Table~\ref{Table1}). Despite the large range of  Deborah numbers in our experiments (from below 1 to above 20), a clear trend  is not apparent in the data.

Instead of the Deborah number, one could  argue that the relevant parameter to interpret the data is the Weissenberg number, $\Wi$, which, instead of  comparing the relaxation time of the fluid with the rotation rate of the swimmer, compares it to the relative rate of deformation in the flow. Hence, we can define this number as
$\Wi =  \dot\gamma \tau$, 
where $\dot\gamma$ is the characteristic shear rate. For a rotating helix, the shear rate scales as $R\omega/\lambda$; therefore, we have $\Wi \sim \left({R}/{\lambda}\right)\De$. 
We plot in Fig.~\ref{results_2} (right)   the normalized mean speed, $U_{NN}/U_N$ as a function of the  Weissenberg number, for all   experiments. Similar to the previous case, the data shows an unidentifiable dependence on $\Wi$. These dimensionless numbers can therefore not be used alone to characterize the changes in swimming speed when viscoelastic effects are present.

\begin{figure}[t]
\noindent\includegraphics[width=0.45\textwidth]{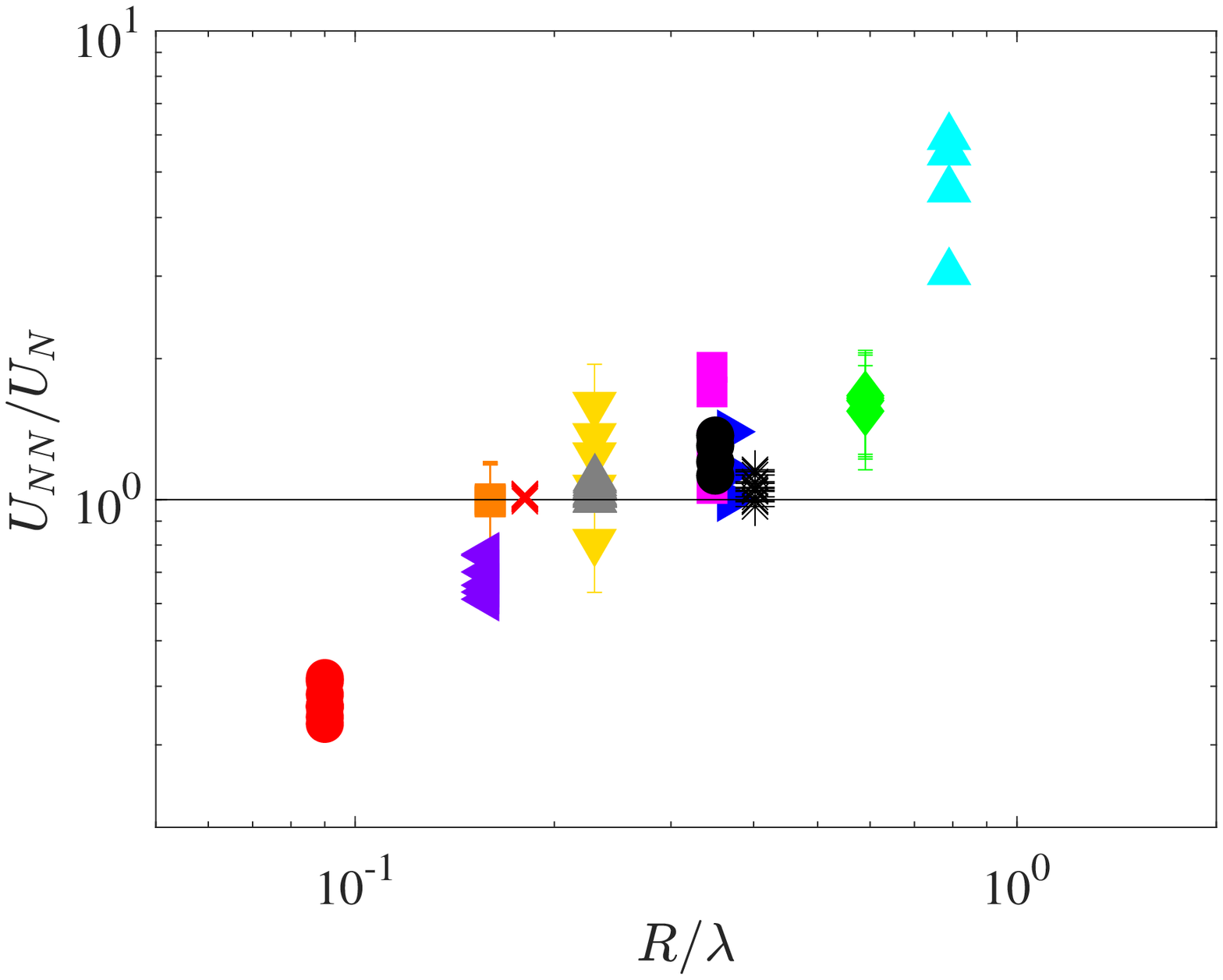}\qquad
\quad\quad
\includegraphics[width=0.45\textwidth]{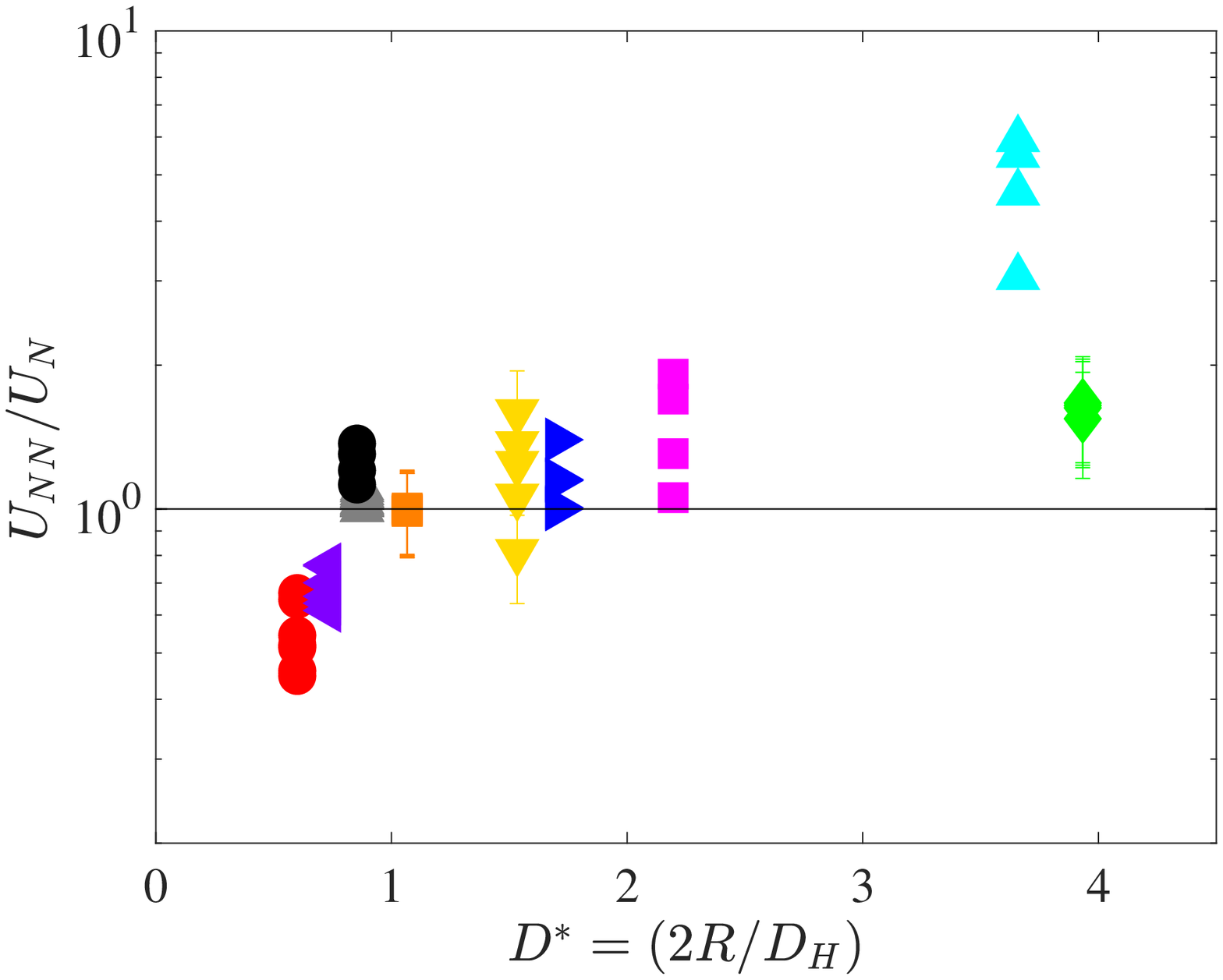}
\caption{(left) Ratio of viscoelastic to Newtonian swimming
speeds, $ U_{NN}/U_N$, as a function of the helix aspect ratio,   $R/\lambda$; the ($\ast$) and ({\color[rgb]{1,0,0}$\times$}) symbols show the data from Ref.~\cite{liu2011} for $R/\lambda=0.40$ and $R/\lambda=0.18$, respectively. (right) Ratio of viscoelastic to Newtonian swimming speeds, $U_{NN}/U_N$, as a function of helix to head diameter ratio $D^*\equiv 2R/D_H$.
All filled experimental symbols follow   Table~\ref{Table1}.}
\label{results_3}
\end{figure}

Contrasting our  data with  the experimental results from Ref.~\cite{liu2011}, we notice that in this work also the dependence of the swimming speeds with $\De$  for helices with different pitch angles did not collapse into a single curve.  The follow-up numerical study in Ref.~\cite{spagnolie2013} showed also  that the ratio $U_{NN}/U_N$ was affected by both the Deborah number and the helix pitch angle. Guided by these studies, we re-plot our  data in Fig.~\ref{results_3} (left) with the swimming speed increase  now shown as   a function of $R/\lambda=\tan\theta/(2\pi)$. Displayed in this manner, we see a remarkably consistent increase of swimming enhancement with $R/\lambda$ (i.e.~with the helix angle, $\theta$) regardless of the value of the  Deborah number. A value of $R/\lambda\approx 0.213$, corresponding to a helix angle of $\theta\approx 53.3^\circ$, appears to mark the transition from a decrease to an increase in swimming speed.  We have also included the data from Ref.~\cite{liu2011}  in  Fig.~\ref{results_3} (left) ($\ast$  and  {\color[rgb]{1,0,0}$\times$} symbols); the small number of data points in that study appear to fit within the  uncertainty of our experiments. Note, however, that the increase in $U_{NN}/U_N$ found by these authors was very modest in comparison to the present data where we obtain increases of up  to a factor of five.

One important aspects of the geometry of the swimmers shown in Table \ref{Table1} is that the size of the head, $D_H$, remains relatively constant for all swimmers; however, to achieve different pitch angles, the size of the helix, $2R$, varies significantly. Therefore, the helix-to-head size ratio, $D^*=2R/D_H$, varies form 0.6 to 3.9. The helix diameter can therefore be smaller, similar or larger that the head diameter. To explore the way in which this change in geometry affects the swimming speed, we show in Fig.~\ref{results_3} (right)  the normalized swimming speed, $U_{NN}/U_N$, as a function of the size ratio $D^*$, for all the experiments conducted in this investigation. Clearly, and similarly to the results in Fig.~\ref{results_3} (left), a correlation can be   be identified; when the head is smaller than the helix, the swimming speed in the viscoelastic fluid is larger that the Newtonian one, and when the head is larger then the opposite happens.
 
\section{Physical interpretation}

How can we explain  theoretically the  influence of viscoelasticity  on the swimming speed ratio,  $U_{NN}/U_N$? While viscoelastic effects are undoubtedly important,  the values of the Deborah or Weissenberg numbers alone are not able to quantify the impact of elastic stresses on the   swimming speed. As   shown above,  both the helix angle, and the helix-to-head size ratio, appear to play a role in the balance between thrust and drag on the swimmer. We consider them both separately in what follows.

\subsection{Local resistive model}
Using the observation, shown  in Fig.~\ref{results_1}, that the swimming speed increases approximately   linearly with  the rotational frequency in all cases, we can first attempt to rationalise the impact of the helical slope  using  resistive-force theory for low-Reynolds number swimmers~\cite{Winet1977}. This is known to be valid in the Newtonian case for slender swimmers, and thus should remain approximately valid at small Deborah numbers in the viscoelastic case. To address the role of the helix angle, we consider the limit of the small swimmer head so that both propulsion and thrust are dominated by the rotating helical tail.

Neglecting the viscous drag on the  head of the swimmer,  the swimming speed of a force-free helix is predicted by  the resistive-force theory framework to be given by
\begin{equation}
\frac{U}{\omega R}= \frac{\left(\xi-1\right)\tan\theta}{1+\xi\tan^2\theta},
\end{equation}
where $\xi=c_\bot/c_\parallel$ is the ratio between the drag coefficient  for local portions of the slender helix moving perpendicularly and parallel to the local tangent~\cite{Winet1977} and $\tan\theta$ is the tangent of the helix angle.  Assuming that a similar local hydrodynamic analysis can be conducted for a viscoelastic Boger fluid at small $\De$, the helix swimming speed would then be 
\begin{equation}
\left(\frac{U}{\omega R}\right)_{NN}= \frac{\left(\xi^{NN}-1\right)\tan\theta}{1+\xi^{NN}\tan^2\theta},
\end{equation}
where $\xi^{NN}=c^{NN}_\bot/c^{NN}_\parallel$ is the drag coefficient ratio for a viscoelastic flow. Hence, assuming that $\xi\approx2$ in the Newtonian case~\cite{Winet1977}, we    obtain
\begin{equation}
\frac{U_{NN}}{U_{N}}=\left(\xi^{NN}-1\right) \left(\frac{1+2\tan^2\theta}{1+\xi^{NN}\tan^2\theta}\right),
\label{eqn:speedratio}
\end{equation}
which can, theoretically, be  smaller or larger that one depending on the the value of $\tan \theta$ and on $\xi^{NN}$.

Our experimental results from Fig.~\ref{results_3} (left) show that $U_{NN}/U_N<1$ for small pitch angles (small $R/\lambda$). This would be consistent with the model in Eq.~\eqref{eqn:speedratio}  in this limit if  $1< \xi^{NN} < 2$, i.e.~for a perpendicular drag that remains larger than the parallel one, but less so than in the Newtonian case. In contrast, for large pitch angles (large $R/\lambda$), the experiments show that $U_{NN}/U_N>1$. This would be consistent with the model in Eq.~\eqref{eqn:speedratio} only if the drag ratio satisfied $\xi^{NN}> 2$.  

There is therefore a contradiction. Of course, such a local resistive-force theory approach   could very well not be valid in a viscoelastic fluid, for example if   nonlocal effects (hydrodynamic interactions) played an important role. Alternatively, if   the local theory was valid,   the ratio of drag  coefficients  $\xi^{NN}$  would have to depend on the  value of the angle  $\theta$, i.e.~the local orientation of the helix relative to the fluid in which it moves. While recent numerical work    reported  that the elastic stresses in the wake of rigid cylinders depend   on the orientation of the cylinder relative to is velocity~\cite{Li2018},  the dependance of  the drag coefficient ratio for different angles in viscoelastic flows have not been reported to date. This resistive-force theory approach does not appear, therefore,  to explain the results from Fig.~\ref{results_3} (left) in a  physically-intuitive way.

\subsection{The snowman effect}
 
 We can, however,  provide a physical mechanism for the change in swimming plotted as in Fig.~\ref{results_3} (right) by turning to past work that addressed the  effect of asymmetry for  rotating swimmers in viscoleastic fluids. These  theoretical~\cite{pak12} and experimental 
studies~\cite{puente2019} showed that a  snowman, i.e.~a dumbbell composed of two spheres of different diameters, could swim in a viscoelastic fluid when  rotating about its symmetry axis. The physical origin of the propulsion lies in the  secondary flows  generated in elastic fluids by normal-stress difference that, for a rotating sphere, lead to fluid flows directed away from the sphere along its rotation axis. A dumbbell made of two spheres of different sizes experiences therefore an imbalance of drag due to these two elastic flows, resulting in swimming.  This viscoelastic propulsion force is directed in the direction from the largest to the smallest sphere~\cite{pak12,puente2019}. 

Our data  in Fig.~\ref{results_3} (right) clearly   indicate that the front-back asymmetry  of the helical swimmers does control the normalized swimming speed. We  propose  therefore that it is the size asymmetry between the head and the tail that leads to an additional snowman-like viscoelastic force affecting the swimming speed. If this mechanism is correct,  and for locomotion that takes place head-first (the case in our experiments), a swimmer with a head smaller than the helix should swim faster due to this viscoelastic {snowman} effect; conversely, if the head is larger than the tail the swimming speed should decrease. This is indeed what we see in our experiments.

In order to be more quantitative, we consider the theoretical expression derived in Ref.~\cite{pak12}, and estimate the additional  viscoelastic force resulting from the difference in size between the head and helix. Assuming as a first approximation that the additional viscoelastic force is generated regardless of the detailed shape of the head or helix, and identifying the diameters of the spheres in Ref.~\cite{pak12} to the diameters of the head and helix in our experiment, the snowman propulsive force predicted theoretically is given by
\begin{equation}
P_{S}=c_S \omega \left(\frac{D_H}{2}\right)^2\, \De \frac{ {D^*}^3({D^*}-1)}{(1+{D^*})^5},
\label{eqn:snowman}
\end{equation}
where $c_S$ is a viscous drag coefficient ($c_S=k \mu$ where $k$ is a dimensionless shape factor), ${D^*}=2R/D_H$ is the rise ratio and $\De=\omega\tau$ is the Deborah number.  Next, we assume for simplicity that the propulsion, $P_{helix}$, and viscous drag on the helix, $D_{helix}$, are not far  from those given by  the Newtonian resistive-force theory, and similarly for the  drag force on the head of the swimmer, $D_{head}$, the steady force balance on the swimmer in a viscoelastic fluid is now given by
\begin{equation}
P_{helix}+P_S=D_{helix}+D_{head}.
\end{equation}
Using the classical expressions for $P_{helix}$, $D_{helix}$ and $D_{head}$ from Ref.~\cite{Rodenborn2013}, and combining them with Eq.~\eqref{eqn:snowman}, we obtain the result
\begin{equation}
U_{NN}=U_N+U_S,
\end{equation}
where $U_N$ is the Newtonian swimming velocity given by
\begin{equation}
U_{N}=\omega R\left(\frac{(\xi-1)\tan\theta}{1+\xi\tan^2 \theta+\xi_oL^*\sec\theta}\right),
\end{equation}
with $L^*=L_H/L$ and $\xi_o=c_H/c_\parallel$ is the normalized head drag coefficient ($c_H$ is the head drag coefficient). The additional snowman speed $U_S$ is given by 
\begin{equation}
U_{S}=\omega R \left(\frac{\xi_S \frac{D_H^*}{2\cos\theta}\De \frac{{D^*}^2({D^*}-1)}{(1+{D^*})^5} }{1+\xi\tan^2 \theta+\xi_oL^*\sec\theta}\right),
\end{equation}
with $\xi_S=c_S/c_\parallel$ and $D_H^*=D_H/L$. 

Using this model,  the additional viscoelastic thrust resulting from the  front-back asymmetry leads to the normalized swimming speed written as a sum
\begin{equation}
\frac{U_{NN}}{U_{\text{N}}}=1 +\frac{U_{\text{S}}}{U_{\text{N}}}
\label{eq:NN_speed},
\end{equation}
where
\begin{equation}
\frac{U_{S}}{U_{\text{N}}}= \frac{\xi_s \De}{2(\xi-1)}\frac{D_H^* }{\sin\theta} \frac{{D^*}^2({D^*}-1)}{(1+{D^*})^5}.
\label{eq:NN_speed2}
\end{equation}
This final expression indicates that the viscoelastic contribution due to the    asymmetry of the swimmer depends on many factors, including the  Deborah number and the size ratio $D^*$. Importantly, the ratio $U_S/U_{\text{N}}$ can be positive or negative depending on the value of $D^*$ relative to one. Since  $\xi>1$, swimmers with $D^*>1$ will swim faster than in the Newtonian fluid while  those with $D^* <1$ will slow down. 

\begin{figure}[t]
\includegraphics[width=0.5\textwidth]{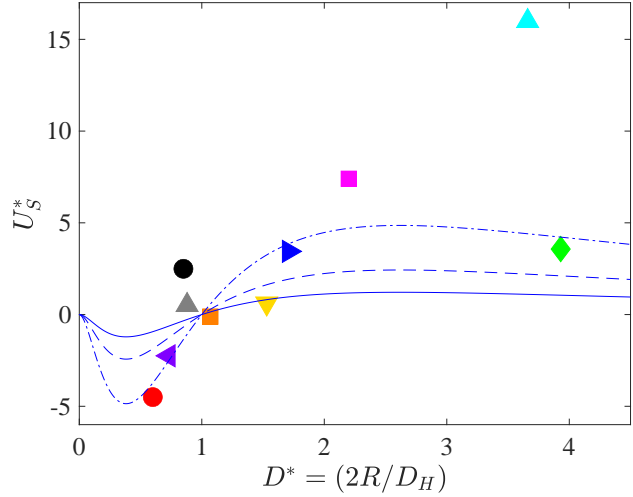}
\caption{Modified extra swimming speed, $U_S^*$ (defined in Eq.~\ref{eq:NN_speed3}), as a function of helix to head diameter ratio $D^*\equiv 2R/D_H$, for  $\De\approx 6.8$. The symbols are  the experimental values while the lines show the theoretical predictions of the model in  Eq.~\eqref{eq:NN_speed3} with values  $\xi_s=20$ (solid line), 40 (dashed line) and 80 (dash-dotted line).}
\label{results_4}
\end{figure}

To show that this snowman model can reproduce   the experimental trend, we   use Eq.~\eqref{eq:NN_speed2} to  define first a modified value of $U_S/U_N$, termed $U_S^*$,  as
 \begin{equation}
U_S^*=\frac{U_{S}}{U_{\text{N}}} \frac{\sin\theta}{D_H^* }= \frac{\xi_s \De}{2(\xi-1)}\frac{{D^*}^2({D^*}-1)}{(1+{D^*})^5},
\label{eq:NN_speed3}
\end{equation}
where both $\sin\theta$ and $D_H^*$ are known quantities in our experiments. 
The value of $U_S^*$ can then be plotted as a function of $D^*$ for given values of $\De$ and $\xi_s$.  To do so, we  extract data from Fig.~\ref{results_2} (left) for an approximately constant value of $\De\approx 6.8$  from which, using Eq.~\eqref{eq:NN_speed}, experimental values of $U_S^*$ can be calculated. We show in 
Fig.~\ref{results_4}  the comparison between the model,  
Eq.~\eqref{eq:NN_speed3} and the experimental values using three possible values for the dimensionless factor $\xi_s$. The model is able to reproduce the experimental trend and shows a clear transition for $U_S/U_N$ from negative to positive values, thus explaining the transition from slower to faster than Newtonian when the helix to tail size ratio goes from smaller to larger than unity.

\section{Conclusion}
In this work we have carried out 
 experiments on the locomotion of free-swimming  magnetically-driven rigid helices  in Newtonian and viscoelastic (Boger) fluids.   We varied the  sizes of the swimmer body and its helical tail and showed that the impact of viscoelasticity    depends critically on the  geometry of the swimmer: it can lead to a large increase of  the swimming speed, a decrease or it can have  approximately no impact. We proposed that the influence of  viscoelasticity on helical propulsion is  controlled by a snowman-like viscoelastic effect, previously reported for dumbbell swimmers, wherein the front-back asymmetry of the swimmer generates a non-Newtonian elastic propulsion force that can either  favor or hinder locomotion. 
 
 The obvious next step in this investigation would be to address a similar question for biological swimmers propelled by helical flagellar filaments.  
 Swimming bacteria such as {\it E.~coli} have a cell body whose width is approximately $D_H\approx 0.88~\mu$m while the diameters of the helical flagella is approximately $2R \approx 0.4~\mu$m. The dimensionless ratio in that case is therefore given by 
 ${D^*}=2R/D_H\approx 0.45$. Since this  is less than one, our results suggest therefore that bacteria self-propeling in similar fluids would have their  swimming speed decreased by elastic stresses.

\section*{Acknowledgements}
V. Angeles is grateful to Conacyt-Mexico for a graduate student scholarship and support. This project has received funding from the European Research Council (ERC) under the European Union's Horizon 2020 research and innovation programme  (grant agreement 682754 to EL).


\end{document}